\documentclass[journal,10pt]{IEEEtran}
\usepackage[cmex10]{amsmath}
\usepackage[utf8]{inputenc}
\usepackage{array}
\usepackage[tight]{subfigure}
\usepackage{cite}
\usepackage{algorithmic}
\usepackage[ruled,vlined]{algorithm2e}
\usepackage{textcomp}
\usepackage{comment}
\usepackage{tikz}
\usepackage{cuted}
\usetikzlibrary{decorations.pathreplacing,decorations.markings,shapes.geometric}
\tikzset{naming/.style={align=center,font=\small}}
\tikzset{antenna/.style={insert path={-- coordinate (ant#1) ++(0,0.25) -- +(135:0.25) + (0,0) -- +(45:0.25)}}}
\tikzset{station/.style={naming,draw,shape=dart,shape border rotate=90, minimum width=10mm, minimum height=10mm,outer sep=0pt,inner sep=3pt}}
\tikzset{mobile/.style={naming,draw,shape=rectangle,minimum width=12mm,minimum height=6mm, outer sep=0pt,inner sep=3pt}}
\tikzset{radiation/.style={{decorate,decoration={expanding waves,angle=90,segment length=4pt}}}}
\newcommand{\BS}[1]{%
\begin{tikzpicture}
\node[station] (base) {S};
\draw[line join=bevel] (base.100) -- (base.80) -- (base.110) -- (base.70) -- (base.north west) -- (base.north east);
\draw[line join=bevel] (base.100) -- (base.70) (base.110) -- (base.north east);
\draw[line cap=rect] ([yshift=0pt]base.north) [antenna=1];
\end{tikzpicture}
}
\tikzset{radiation/.style={{decorate,decoration={expanding waves,angle=90,segment length=4pt}}}}
\newcommand{\R}[1]{%
\begin{tikzpicture}[every node/.append style]
 \draw[semithick] (0,1) -- (1,4);
            \draw[semithick] (3,1) -- (2,4);
            \draw[semithick] (0,1) arc (180:0:1.5 and -0.5);
            \node[inner sep=2pt] (circ) at (1.5,5.5) {};
            \draw[semithick] (1.5,5.5) circle(4pt);
            \draw[semithick] (1.5,5.5cm-4pt) -- (1.5,4);
            \draw[semithick] (1.5,4) ellipse (0.5 and 0.150);
            \draw[semithick,radiation,decoration={angle=45}] (1.5cm+4pt,5.5) -- +(0:2);
            \draw[semithick,radiation,decoration={angle=45}] (1.5cm-4pt,5.5) -- +(180:2);
\end{tikzpicture}
}
\newcommand{\U}{%
\begin{tikzpicture}[every node/.append style]
\begin{scope}[line join=round,looseness=0.25, line cap=round]
\clip [preaction={left color=blue!10, right color=blue!30}] 
  (1/2,-1) to [bend left] (0,10)
  to [bend left] ++(1,1) -- ++(0,2)
  arc (180:0:3/4 and 1/2) -- ++(0,-2)
  to [bend left]  ++(5,-2) coordinate (A) to [bend left] ++(-1/2,-11)
  to [bend left] ++(-1,-1) to [bend left] cycle;
\path [left color=blue!30, right color=blue!50]
  (A) to [bend left] ++(0,-11) to[bend left] ++(-3/2,-2)
  -- ++(0,12);
\path [fill=blue!20, draw=white, line width=0.25cm]
  (0,10) to [bend left] ++(1,1) -- ++(0,2)
  arc (180:0:3/4 and 1/2) -- ++(0,-2)
  to [bend left]  (A) to [bend left] ++(-3/2,-5/4)
  to [bend right] cycle;
\draw [line width=0.25cm, fill=white]
  (9/8,21/2) arc (180:360:5/8 and 3/8) --
  ++(0,2.5) arc (0:180:5/8 and 3/8) -- cycle;
\draw [line width=0.25cm, fill=white]
  (9/8,13) arc (180:360:5/8 and 3/8);
\fill [white, shift=(225:0.5)] 
  (1,17/2) to [bend left] ++(4,-7/4)
  to [bend left] ++(0,-7/2) to [bend left] ++(-4, 6/4)
  to [bend left] cycle;
\fill [black, shift=(225:0.25)] 
  (1,17/2) to [bend left] ++(4,-7/4)
  to [bend left] ++(0,-7/2) to [bend left] ++(-4, 6/4)
  to [bend left] cycle;
\shade [inner color=white, outer color=cyan!20] 
  (1,17/2) to [bend left] ++(4,-7/4)
  to [bend left] ++(0,-7/2) to [bend left] ++(-4, 6/4)
  to [bend left] cycle;
\draw [line width=0.25cm, shorten <=-0.5cm, shorten >=-0.5cm]
  (1,17/2) to [bend left] ++(4,-7/4)
  to [bend left] ++(0,-7/2);
\foreach \i in {1,2,3}
  \foreach \j in {1,2,3}
     \draw [line width=0.125cm, top color=gray, 
       bottom color=white, yslant=-3/8, rounded corners=1cm/8] 
         (\i*3/2-1/2,-\j+4) rectangle ++(1,3/4);
\draw [line width=0.125cm, top color=gray, bottom color=white, yslant=-3/8] 
  (3,0) ellipse [x radius=1/2, y radius=1/3];
\draw [line width=0.125cm, top color=gray, bottom color=white, yslant=-3/8] 
  (3,19/2) ellipse [x radius=1, y radius=1/4];
\draw [line width=0.5cm] 
  (1/2,-1) to [bend left] (0,10)
  to [bend left] ++(1,1) -- ++(0,2)
  arc (180:0:3/4 and 1/2) -- ++(0,-2)
  to [bend left]  ++(5,-2) to [bend left] ++(-1/2,-11)
  to [bend left] ++(-1,-1) to [bend left] cycle;
\end{scope}
\end{tikzpicture}
}
\usetikzlibrary{positioning}
\usepackage{marvosym}
\usepackage{footnote}
\usepackage{epstopdf}
\usepackage{mathtools,amssymb}

\makeatletter
\renewcommand{\fnum@figure}{Fig. \thefigure}
\makeatother
\usepackage{cite,amsmath,amsthm,amssymb,float,latexsym,epsfig,subfigure,epstopdf,array}
\newtheorem{theorem}{Theorem}
\newtheorem{lemma}[theorem]{Lemma}
\usepackage[font=small,labelfont=bf,labelsep=space]{caption}
\usepackage{float} 
\usepackage{xcolor} 
\usepackage[margin=1in]{geometry}
\usepackage{multirow,booktabs}
\newcommand{\MeijerG}[7]{G \begin{smallmatrix} #1 & #2 \\ #3 & #4 \end{smallmatrix} \left( \begin{smallmatrix} #5 \\ #6 \end{smallmatrix} \middle\vert #7 \right) }
\begin{document}
	\title{On Performance of Energy Harvested Cooperative NOMA Under Imperfect CSI and Imperfect SIC }
		\author{Author 1, Author 2 Author 3 }
	\author{Shubham Bisen, Parvez Shaik and Vimal Bhatia
	\thanks{Shubham Bisen, Parvez Shaik and Vimal Bhatia are with the Discipline of Electrical Engineering, Indian Institute of Technology Indore, Indore 453552, India (e-mail: \{phd1901202001; phd1601202003;  vbhatia\}@iiti.ac.in.)}
}
\maketitle
\begin{abstract}
With the advent of 5G and the need for energy-efficient massively connected wireless networks, in this work, we consider an energy harvesting (EH) based multi-relay downlink cooperative non-orthogonal multiple access (NOMA) system with practical constraints. The base station serves NOMA users with the help of decode-and-forward based multiple EH relays, where relays harvest the energy from the base station's radio frequency. A relay is selected from the multiple K-relay by using a partial relay selection protocol.  The system is considered to operate in half-duplex mode over a generalized independent and identical Nakagami$-m$ fading channel. The closed-form expression of outage probability and ergodic rate are derived for users, under the assumption of imperfect channel state information (CSI) and imperfect successive interference cancellation (SIC) at the receiver node. Expression of outage probability and ergodic rate for both users under the assumption of perfect CSI  and perfect SIC are also presented. Further, the asymptotic expression for the outage probability is also shown. The derived analytical expressions are verified through Monte-Carlo simulations.
\end{abstract}
	\vspace{-2mm}
\IEEEpeerreviewmaketitle
	\section{Introduction}
$5^{th}$ generation (5G) wireless communication standard is drafted by the third generation partnership project to meet the growing demands of multi-rates data applications. Non-orthogonal multiple access (NOMA) is one of the leading beyond 5G technologies and is considered a prominent and promising solution with superior spectrum efficiency and massive connectivity \cite{ding2017survey, swami2018cooperation,8114722}.
Recently, cooperative relaying and the NOMA system have gained research attention due to its significant and prominent advantages in improving the coverage and reliability. \cite{7230246,ding2015cooperative,NOMAmultiuser,RelayselectionNOMA2016,optimalRS2018,Perforrelayselectin,PSRIETNOMA,justin_vfd}. In \cite{ding2015cooperative}, authors investigated cooperative-NOMA with amplify-and-forward (AF) and decode-and-forward (DF) relaying and derived the outage probability and asymptotic outage probability expressions. Considering the fact that multiple relays provide a significant performance gain in cooperative networks, special attention has been given to the cooperative NOMA with multiple relays and analysis of relay selection techniques \cite{RelayselectionNOMA2016,optimalRS2018,Perforrelayselectin,PSRIETNOMA}. In \cite{RelayselectionNOMA2016}, the authors have provided a two-stage max-min relay selection scheme for the NOMA system with fixed power allocation at relays and different quality of service (QoS) requirements at users.   In \cite{RelayselectionNOMA2016}, the users are selected based on different QoS requirements, whereas in \cite{optimalRS2018},  authors have considered the users according to their channel condition, with fixed and adaptive power allocation at the relay node. In \cite{Perforrelayselectin}, the outage probability of cooperative NOMA with multiple half-duplex DF relays is analyzed. In \cite{PSRIETNOMA}, the outage probability and sum rate of cooperative NOMA with multiple half-duplex AF relays were studied, considering partial relay selection (PRS).
Energy harvesting (EH) from radio frequency signals is considered a viable solution to provide additional lifespan to energy-constrained nodes. Hence, EH based NOMA systems have gained research attention to meet the needs of 5G and beyond communications.  In  \cite{SWIPTNOMA,EHOP,IETEH,PerEHOE,jointharcsi,Mvaez},  the performance of cooperative NOMA has been studied with EH relaying. A cooperative NOMA with simultaneous wireless information and power transfer is studied in \cite{SWIPTNOMA}, where nearby users acting as EH relays assist the far away NOMA users.  In \cite{EHOP}, authors analyzed the performance of cooperative NOMA network with EH relaying over Rayleigh fading. In \cite{IETEH}, a multi-user NOMA system is analyzed with an EH powered relay node. In \cite{PerEHOE}, the authors evaluated the NOMA system's performance with multiple EH relaying and derived the closed-form expressions of outage probability and ergodic capacity over the Rayleigh fading channel. In \cite{jointharcsi}, the authors analyzed the NOMA system's outage probability with multiple EH AF relays with imperfect CSI and hardware impairment. \\
 In the literature, most of the work assumes perfect knowledge of the channel state information (CSI) at the receiver, which is too idealistic for a practical system. However, in practice, knowledge of perfect CSI is unavailable at the receivers, which leads to channel estimation error (CEE). CEE significantly deteriorates the performance of the system \cite{ICSINOMA,PshkICSI,jointharcsi}. In \cite{ICSINOMA},  a closed-form expression of outage probability is derived of a downlink relay aided NOMA system with multiple users over Nakagami-$m$ fading with imperfect CSI. 
 \vspace{-1mm}
\subsubsection*{Contributions} 
To the best of the author's knowledge, the NOMA system's performance with decode-and-forward (DF) based multiple EH relays over  Nakagami$-m$ fading channels has not been considered. Further, a detailed study on the impact of practical constraint, imperfect CSI at receiver nodes, and imperfect SIC is not available in the literature. Nakagami-$m$ fading is considered due to its generalization to a variety of realistic fading channels, which includes Rayleigh fading channel with $m$ = 1, recently for THz channels with $m$ = 3 \cite{THZ}, and also for the unmanned aerial vehicle (UAV) \cite{UAV}.  The main contributions of this work are:
\vspace{-1mm}
\begin{itemize}
  \item For the first time in EH based multi-relay  NOMA system, we consider the practical case of imperfect CSI at receiver nodes and imperfect SIC, and its impact on the system is analyzed.
  \item We present a DF based cooperative multiple EH relay based NOMA system employing relay selection and analyze its performance in terms of outage probability by deriving its closed-from expression from the novel end-to-end (e2e) SNR of the considered system for both users for both perfect and imperfect cases.
  \item Further, we analyzed the system performance at high SNR by deriving the expression of asymptotic outage probability and useful insights are drawn.
  \item We optimized the EH time fraction parameter  ($\alpha$) to maximize the system throughput. The optimization problem is solved by adopting the particle swarm optimization (PSO) under both perfect and imperfect CSI/SIC conditions.
 \item Rate analysis of the system is performed by deriving the closed-form expression of ergodic rate for both users and the impact of imperfect CSI and SIC is analyzed.
   \end{itemize}
The rest of the paper is organized as follows:  the system model is introduced in Section II.  Outage probability analysis is presented in Section III. Asymptotic outage probability analysis is presented in Section IV. Ergodic rate is analyzed in Section V. Numerical and simulation results are discussed in Section VI. Finally, in Section VII, conclusions are drawn. \\
	\textit{Notation}:         %
		%
		$\text{$\Gamma$}(\cdot) $ is gamma function,   $\MeijerG{m}{n}{p}{q}{a_1,\ldots,a_p}{b_1,\ldots,b_p}{x}$ is MeigerG function.
		Square of the norm is denoted by $|\cdot|^2$.
		%
		Complex Gaussian distribution with mean 0, variance $\sigma^2$ is denoted by $\mathcal{CN}(0,\sigma^2)$  and 
		modified Bessel function of second kind of order $n$ is denoted by  $K_{n}(\cdot)$. 
		Expectation operator is denoted by $\text{E}(\cdot)$. 
		%
		%
		%
		Gaussian random variable (RV) with mean $\mu$ and variance $\sigma^2$ is represented as $\mathcal{N}({\mu, \sigma^2})$. 
		Probability density function (PDF) and cumulative distribution function (CDF) are given by $f(\cdot)$ and $F(\cdot)$, respectively. 
		Factorial is denoted by $(\cdot)!$.
		%
		%
		
\vspace{-5mm}
 \section{System Model} 
\begin{figure}
    \centering
\begin{tikzpicture}[scale=0.5]
    \node[align=center,scale=0.12, label={$R_1$}] (r1) {\R{$R_1$}};
    \node[align=center,below=0.4cm of r1, scale=0.12,label={$R_2$}] (r2) {\R {$R_2$}};
    \node[fill=black,circle,inner sep=1pt,below=0.2cm of r2](c1){};
    \begin{scope}[node distance=0.2cm,inner sep=1pt]
        \node[fill=black,circle,below=0.2cm of c1](c2){};
        \node[fill=black,circle,below=0.2cm of c2](c3){};
    \end{scope}
    \node[scale=0.12,align=center,below=0.3cm of c3,label={$R_K$}] (rK) {\R{$R_K$}};
    est] at ([yshift=-0.2cm]uK.east) {};
    \node[right=3cm of r2,right=1.5cm of r1,scale=0.09,,label={$U_1$}](U1) {\U};
    \node[right=2cm of r2,right=3cm of rK, scale=0.09,label={$U_2$}] (u2) {\U};
    \node[left=2cm of c1,label=Source] (Bs){\BS{1}}; 
   \draw[-latex] ([xshift=0.1cm]r2.east) to node[midway,above,sloped]{\textbf{$h_{r_k,2}$}} ([xshift=-0.0cm,yshift=0.0cm]u2.north);
    \draw[-latex] ([xshift=0.1cm]r2.east) to
    node[midway,above,sloped]{\textbf{$h_{r_k,1}$}}
    ([xshift=-0.2cm,yshift=0cm]U1.north);
    \draw[-latex] ([xshift=-0.0cm]Bs.north) to node[midway,above,sloped]{\textbf{$h_{s,r_1}$}} ([xshift=-0.1cm,yshift=-0.01cm]r1.west);
    \draw[-latex] ([xshift=0.0cm]Bs.north) to node[midway,above,sloped]{\textbf{$h_{s,r_2}$}} ([xshift=-0.1cm,yshift=-0.1cm]r2.west);
    \draw[-latex] ([xshift=0.0cm]Bs.north) to node[midway,above,sloped]{\textbf{$h_{s,r_K}$}} ([xshift=-0.1cm,yshift=-0.1cm]rK.west);
\end{tikzpicture}
  \caption{NOMA system with multiple EH relays}
    \label{fig:sysmom}
\end{figure}
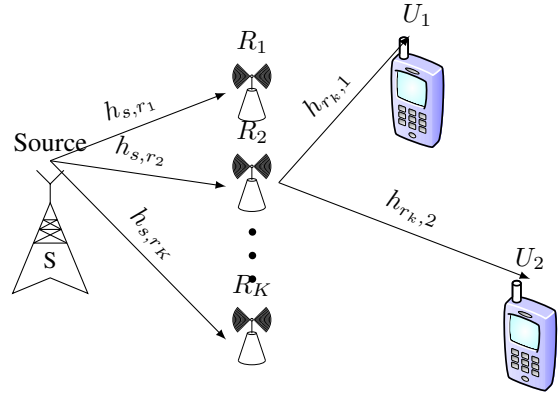
 We consider a downlink cooperative NOMA system with EH-multiple relays as  shown in \figurename{ \ref{fig:sysmom}}, where a transmitter $(S)$, i.e., the base station (BS) transmits messages to the downlink users, i.e., $U_1$ and $U_2$, with the help of cooperative relaying \cite{RelayselectionNOMA2016,optimalRS2018}. Both users are ordered according to their channel condition. It is assumed that the weak users ($U_2$) is with poor channel condition and strong users ($U_1$) is with good channel condition. To facilitate cooperative relaying, it is also assumed that there are K numbers of relays in the system with DF relaying.  Further, it is assumed that each  node is equipped with a single antenna and operates in half-duplex mode. All the channel links are considered to be independent and identically distributed (i.i.d), which are modeled by Nakagami$-m$ fading, the channel coefficients corresponding to links are represented as $h_{i,j}$, $i\in(s,r_k)$ and $j\in(r_k,1,2)$, where the subscript $s,r_k,1$ and $2$ represent BS, $R_k$, $U_1$ and  $U_2$ respectively. The $|h_{i,j}|$ is assumed to be Nakagami$-m$ with shape parameter $m_{i,j}$ and variance  $E[|h_{i,j}|^2]=\lambda_{i,j}$.  Under imperfect CSI, according to minimum mean squared error (MMSE) estimation \cite{SMKay,R2,Do_icsi}, 
 \begin{equation}
     h_{i,j}=\hat{h}_{i,j}+\epsilon_{i,j},  
 \end{equation}
 where $h_{i,j}$ is actual channel and  $\hat{h}_{i,j}$ is the estimate of the channel $h_{i,j}$, where  $h_{i,j}$ and $\hat{h}_{i,j}$ are jointly ergodic and stationary  Gaussian process \cite{andriya}. The  $\epsilon_{s,r}$ is the CEE, which is assumed to be complex normal with mean zero and variance $\sigma^2_{e}$
 \cite{vcmleung}. Assuming that $\hat{h}_{i,j}$ and $\epsilon_{i,j}$ are  independent\footnote{The independence of $\hat{h}_{i,j}$  and $\epsilon_{i,j}$ was not required, only that they were uncorrelated \cite{SMKay}.}, thus the estimated channel variance is given as  $\hat{\lambda}_{i,j}=\lambda_{i,j}-\sigma_{e}^2$ \cite{LLscary}. \\
 In this work, the partial relay selection scheme (PRS) is considered for the selection of relay node, where the BS selects the relay which provides the best instantaneous channel gain between the BS and relay \cite{AFPRS}. The BS continuously monitors the quality of the links between the BS and relays and based on this information, the source selects the best link. The PRS scheme to select the best source to relay link is expressed as $|\hat{h}_{s,r}|^2=\arg \max\limits_{k=1, 2,...,K}|\hat{h}_{s,r_k}|^2$.\\
Furthermore, relays harvest the energy from the source transmitted signal. In this paper, we considered the time switching (TS) protocol to harvest energy for $\alpha T$ block time, where $\alpha$ $(0\leq\alpha\leq1)$ is the fraction of block time over which relay harvest the energy from a source transmitted signal. The remaining $(1-\alpha)T$ block time is assigned for the information transfer. The information transfer completes in two blocks, the first  half of  time block $(1-\alpha)\frac{T}{2}$ is assigned for source to relay transmission, and the remaining half     $(1-\alpha)\frac{T}{2}$ for relay to users transmission. Thus, the harvested energy for $\alpha T$ time is given by \cite{ParSWIPTNCC} $E_H=P_s\mu |\hat{h}_{s,r}|^2\alpha T$, where $ 0\leq\mu\leq1$ denotes energy conversion efficiency. The transmit power at $R_k$ is expressed as $P_r=\frac{E_H}{(1-\alpha)T/2}=\frac{2 P_s\mu |\hat{h}_{s,r}|^2\alpha  }{1-\alpha}.$
 Initially, BS transmits the NOMA signal for users by performing power domain multiplexing and superposition coding. The transmitted signal is given by $x_s=\sum_{i=1}^2\sqrt{a_i}{x_i}$, where, ${x_i}$ denotes complex modulated symbols with unit energy for $U_i$ (i.e., $\text{E}\left\{|x_i|^2\right\}=1$). Further, $a_i$ is the power allocation coefficients for $x_i$ with $\sum_{i=1}^2a_i=1$ and $a_i>0$. The received signal at $R_k$ is given by
	$y_{r_{k}}=(\hat{h}_{s,r}+ \epsilon_{s,r_k})\sqrt{P_s}x_s+z_{s,r_k},$
	where, $P_s$ indicates the total transmit power at BS, $z_{s,r}$ represents additive white Gaussian noise (AWGN) with zero mean and variance $\sigma^2_0$. As DF transmission protocol is employed at the relay, $R_k$ has to first decode both $x_1$ and $x_2$ before transmitting. 
	The signal-to-interference noise ratio (SINR) at $R_k$ to decode $x_2$ in the presence of imperfect CSI is given by
	\begin{align}\label{rf}
	\gamma_{r_k,2}=\frac{|\hat{h}_{s,r}|^2P_s a_2}{|\hat{h}_{s,r}|^2 P_s a_1+P_s\sigma^2_{e}+\sigma^2_0}.
	\end{align} 
	  According to SIC principle, $x_1$ is decoded by removing  $x_2$ from $y_{r_k}$, the SIC is perfect, $x_2$ will be completely removed. Otherwise decoding of $x_1$ will be carried out in the presence of residual interference due to imperfect SIC. Thus SINR in the presence of imperfect CSI and imperfect SIC at $R_k$ to decode $x_1$ is given by
	\begin{align}\label{rn}
	\gamma_{r_k,1}=\frac{|\hat{h}_{s,r}|^2P_sa_1}{|\hat{h}_{s,r}|^2 \beta P_s a_2 +P_s a_1\sigma^2_{e}+P_s \beta a_2\sigma^2_{e}  +\sigma^2_0}. 
	\end{align} 
	where, $\beta$ represents the residual interference due to imperfect SIC, $0\leq\beta\leq1$, and $\beta=0$ refer to perfect SIC.  
	After decoding ${x_2}$ and $x_1$, $R_k$ will re-encode the information bit using superposition coding. Thus the transmitted signal by $R_k$ is given by $x_r=\sum_{i=1}^2\sqrt{a_i}{x_i}$. The received signal at $U_1$ and $U_2$ is given as $y_l=(\hat{h}_{r_k,l}+\epsilon_{r_k,l})\sqrt{P_r}x_r+ z_{r_k,l},\; l=1, 2.$ Where $P_r$ is harvested power at $R_k$, $z_{r_k,l}$  represent AWGN with zero mean and variance $\sigma^2_0$. Thereafter, $U_1$ will performs SIC to obtain its own symbol $x_1$. Thus, the SINR at $U_1$ to decode  $x_2$ and  $x_1$ under imperfect CSI and imperfect SCI are given as
		\vspace{-2mm}
	\begin{align}\label{nf}
	\gamma_{1,2}=\frac{|\hat{h}_{r_k,1}|^2P_r a_2}{|\hat{h}_{r_k,1}|^2 P_r a_1+P_r\sigma^2_{e}+\sigma^2_0},
	\end{align} 
		\vspace{-3mm}
	\begin{align}\label{nn}
	\gamma_{1,1}=\frac{|\hat{h}_{r_k,1}|^2P_ra_1}{|\hat{h}_{r_k,1}|^2 \beta P_r a_2 +P_r a_1\sigma^2_{e}+P_r \beta a_1\sigma^2_{e}  +\sigma^2_0}. 
	\end{align} 
   The SINR to decode $x_2$ at $U_2$ under imperfect CSI is given by  
   	\vspace{-3mm}
	\begin{align}\label{eqC}
	\gamma_{2,2}=\frac{|\hat{h}_{r_k,2}|^2P_r a_2}{|\hat{h}_{r_k,2}|^2 P_r a_1+P_r\sigma^2_{e}+\sigma^2_0}.
	\end{align} 
	\section{Outage Probability Analysis}
	In this section, we derive the outage probability expression  for both NOMA users. Outage probability gives the achievable maximum rate for error-free transmission \cite{ieeeaccparvez}. The closed-form outage probability expressions for $U_1$ and $U_2$ under imperfect CSI/ SCI and perfect CSI/SIC are obtained in the following subsections.
		\vspace{-4mm}
	\subsection{Outage probability of $U_1$ under  imperfect CSI and imperfect SIC }
	$U_1$ is said to be in outage, when $U_1$ fails to detect either of the two symbols $x_1$ and $x_2$ at $R_k$ and $U_1$. Therefore the outage probability of $U_1$ is defined as
	%
	\begin{align}\label{Pn}
	 P_{out,1} =& \Pr\big\{\gamma_{r_k,2}  < \gamma_{th2}, \,\nonumber \gamma _{r_k,1}<\gamma_{th1},\,\\& \gamma _{1,2} < \gamma_{th2},\, \gamma _{1,1}<\gamma_{th1}  \big\},
	\end{align} 
	where, $\gamma_{th1}$ and $\gamma_{th2}$ are predefined SINR threshold.  $\gamma_{th1}$  and $\gamma_{th2}$ can be represented as $\gamma_{th1}=2^{\frac{2r_1}{1-\alpha}}-1$ and $\gamma_{th2}=2^{\frac{2r_2}{1-\alpha}}-1$, where, $r_1$ and $r_2$ are  the  desired  target rates. \vspace{-3mm}
	\begin{lemma} When $\gamma_{th2}<\frac{a_2}{a_1}$, $\gamma_{th1}<\frac{a_1}{\beta a_2}$, $\lambda_2>\lambda_1$ and $\Omega_1>\Omega_2$. The outage probability of $U_1$ under imperfect CSI and imperfect SIC is defined in \eqref{A}
		\end{lemma}
	 In \eqref{A},  $A_1=\frac{K}{(m_{s,r_k}-1)!}\sum_{n=0}^{m_{r_k,1}-1}\sum_{j=0}^{n}\frac{1}{n!}\hat{\beta}_{r_k,1}^n\binom{n}{j}$, $A_2=\sum_{k=1}^{K-1}\bigcup_k\Xi_{1,k}\Xi_{2,k}\left(-1\right)^k \binom{K-1}{k}, \,	 \kappa_1=m_{s,r_k}-j-l_1, \, \kappa_2=m_{s,r_k}-j-l_2+\Bar{i},\, \lambda_1=\frac{\gamma_{th2} \sigma_0^2\left(1-\alpha\right)}{\left(a_2-a_1\gamma_{th2}\right)\alpha\mu2\rho}, \,$
	$	\lambda_2=\frac{\gamma_{th1} \left(1-\alpha\right)}{\left(a_1-a_2\beta\gamma_{th1}\right)\alpha\mu2\rho}, \, \Omega_1=\frac{\gamma_{th2 }\sigma_{e}^2}{a_2-a_1\gamma_{th2}},\, \Omega_2=\frac{\gamma_{th2}\left(a_1+a_2\beta\right) \sigma_e^2}{a_1-a_2\beta\gamma_{th2}}, \,\gamma=\max\left[\psi,\left(\frac{\lambda_2-\lambda_1}{\Omega_1-\Omega_2}\right)\right],	\, \rho=\frac{P_s}{\sigma^2_0},	\, \hat{\beta}_{r_k,1}=\frac{m_{r_k,1}}{\hat{\lambda}_{r_k,1}},\,\hat{\beta}_{s,r_k}=\frac{m_{s,r_k}}{\hat{\lambda}_{s,r_k}},\,  \psi=\max\left[\Delta_1, \frac{\gamma_{th1}\left(  \sigma_e^2\left(a_1+a_2\beta\right)+(1/\rho)\right)}{\left(a_1-a_2\beta\gamma_{th1}\right)} \right],\,\Delta_1=\frac{\gamma_{th2}\left(\sigma_e^2+(1/\rho)\right)}{\left(a_2-a_1\gamma_{th2}\right)},\,\iota_1=\sum_{l_1=0}^{N_t}\frac{(-1)^{l_1}}{l_1!}, \, \iota_2=\sum_{l_2=0}^{N_t}\frac{(-1)^{l_2}}{l_2!},  $\\\\\\\\

		\begin{strip}
		\small
	\begin{align}\label{A}
	     P_{out,1}=&1-\Bigg[e^{-\Omega_1\hat{\beta}_{r_k,1}}A_1\Omega_1^{n-j}\lambda_1^j\Big[\iota_1 {\left(\hat{\beta}_{r_k,1}\lambda_1\right)}^{l_1} \nonumber\hat{\beta}_{s,r_k}^{(j+l_1)}\Gamma\left(\kappa_1,\hat{\beta}_{s,r_k}\gamma\right)+A_2\nonumber \left(\hat{\beta}_{r_k,1}\lambda_1\right)^{l_2} \hat{\beta}_{s,r_k}^{(j+l_2-\Bar{i})}\Gamma\left(\kappa_2,\hat{\beta}_{s,r_k}(k+1)\gamma\right)\\&\iota_2\left(k+1\right)^{-\kappa_2}\Big] +e^{-\Omega_2 \hat{\beta}_{r_k,1}}A_1\Omega_2^{n-j}\lambda_1^j\bigg[\iota_1 \left( \hat{\beta}_{r_k,1}\lambda_1\right)^{l_1}\nonumber \nonumber\hat{\beta}_{s,r_k}^{(j+l_1)}\Big[\Gamma\left(\kappa_1,\hat{\beta}_{s,r_k}\psi\right)-\Gamma \Big(\kappa_1,\,\frac{\hat{\beta}_{s,r_k}(\lambda_2-\lambda_1)}{\Omega_1-\Omega_2}\Big)\Big] \nonumber\\&+A_2
	   \iota_2 \left( \hat{\beta}_{r_k,1}\lambda_1\right)^{l_2} \left(k+1\right)^{-\kappa_2} \hat{\beta}_{s,r_k}^{(j+l_2-\Bar{i})} \Big[\Gamma\Big(\kappa_2,(k+1)\hat{\beta}_{s,r_k}\psi\Big)-\Gamma\Big(\kappa_2,\frac{(k+1)\hat{\beta}_{s,r_k}(\lambda_2-\lambda_1)}{(\Omega_1-\Omega_2)}\Big)\Big]\bigg]\Bigg].
	\end{align}
	  \noindent\rule{\textwidth}{0.4pt}
	\end{strip}
	$   \bigcup_k=\sum_{i_1}^{k}\sum_{i_2}^{k-i_2}...\sum_{i_{m_{s,r_k}}}^{k-i_1'''i_{m_{s,r_k}}-2},\, 
	\\ \Xi_{1,k}= \binom{k}{i_1} . . . \binom{k-i_1-...i_{m_{s,r_k}}}{i_{m_{s,r_k}}},\Xi_{2,k}=\prod_{n=0}^{m_{s,r_k}-2}\left( \frac{\hat{\beta}_{s,r_k}^n}{n!} \right)^{i_{n}+1}\left( \frac{\hat{\beta}_{s,r_k}^{m_{s,r_k}-1}}{(m_{s,r_k}-1)!} \right)^{k-i_1-...i_{m_s,{r_k}}-1} 
   $ and $\Bar{i}= (m_{s,r_k}-1)(k-i_1)-(m_{s,r_k}-2)i_2-(m_{s,r_k}-3)i_3-...i_{m_{s,r_k}-1}$. Further, when  $\gamma_{th2}>\frac{a_2}{a_1}$, $\gamma_{th1}>\frac{a_1}{\beta a_2}$, $P_{out,1}=1$.
   \textit{Proof:} See Appendix A	
   \vspace{-3mm}
\subsection{Outage probability of $U_2$ under imperfect CSI}
	$U_2$ is said to be in outage, when $U_2$ fails to detect  $x_2$ at $R_k$ and $U_2$. \vspace{-2mm}
		\begin{align}\label{eq1}
	 P_{out,2} =& \Pr\big\{\gamma _{r_k,2} < \gamma_{th2}, \, \gamma _{2,2}  < \gamma_{th2}\big\},
	\end{align} \vspace{-3mm}
	\begin{lemma} The outage probability of $U_2$ under imperfect CSI is given as\vspace{-3mm}
	\begin{align}\label{P_F_I}
	\small
	    P_{out,2}=&1-e^{-\Omega_1\hat{\beta}_{r_k,2} }A_3\Omega_1^{n-j}\lambda_1^j\nonumber
	    \Big[\iota_1 \left(\hat{\beta}_{r_k,2} \lambda_1\right)^{l_1} \nonumber
	    \nonumber\hat{\beta}_{s,r_k} ^{(j+l_1)}
	  \\&  \Gamma\left(\kappa_1,\,\hat{\beta}_{s,r_k} \Delta_1\right)
	   \nonumber
	   +A_2 \iota_2 \left(\hat{\beta}_{r_k,2} \lambda_1\right)^{l_2}\nonumber \hat{\beta}_{s,r_k}^{(j+l_2-\Bar{i})}
	    \\& \left(k+1\right)^{-\kappa_2}
	   \Gamma\left(\kappa_2,\,(k+1)\hat{\beta}_{s,r_k} \Delta_1\right)\Big],
	\end{align}
	\end{lemma}
	 where $A_3=\frac{K}{(m_{s,r_k}-1)!}\sum_{n=0}^{m_{r_k,2}-1}\sum_{j=0}^{n}\frac{1}{n!}\left(\frac{m_{r_k,2}}{\lambda_{r_k,2}}\right)^n \\ \binom{n}{j}$. Further when  $\gamma_{th2}>\frac{a_2}{a_1}$,  $P_{out,2}=1$.\\
	 \textit{Proof:} See Appendix B
	 \vspace{-3mm}
	\subsection{Outage probability of $U_1$ under perfect CSI and perfect SIC}\vspace{-1mm}
	The outage probability of $U_1$ under perfect CSI and perfect SIC is obtained by considering the perfect channel estimation and perfect SIC at all nodes. The $|\hat{h}_{i,j}|^2=|h_{i,j}|^2$ i.e., the CEE $\epsilon_{i,j}=0$ and residual interference due to imperfect SIC $\beta=0$. 
	\begin{lemma}The outage probability of $U_1$ under perfect CSI/SIC is expressed as\vspace{-3mm}
	\begin{align}\label{p_N}
	\small
	    P_{out,1}^{P}=&1-B_1
	    \left(\beta_{r_k,1} \Delta_2\right)^n
	    \Big[\iota_1 \left(\beta_{r_k,1} \Delta_2\right)^{l_1} \nonumber\\& \nonumber\Gamma\left(\kappa_3,\,\beta_{s,r_k} \Delta_3\right)
	   \nonumber\beta_{s,r_k}^{(n+l_1)} +A_2
	    \nonumber\iota_2 \left(\beta_{r_k,1}\Delta_2\right)^{l_2} \\& \left(k+1\right)^{-\kappa_4}\beta_{s,r_k}^{(n+l_2-\Bar{i})}\Gamma\left(\kappa_4,\,(k+1)\beta_{s,r_k}\Delta_3\right)\bigg],
	\end{align}
	\end{lemma}
	\vspace{-3mm}
		where $B_1=\frac{K}{(m_{s,r_k}-1)!}\sum_{n=0}^{m_{r_k,1}-1}\frac{1}{n!},\,\kappa_3=m_{s,r_k}-n-l_1,$ $ \, \kappa_4=m_{s,r_k}-n-l_2+\Bar{i},\,$ $\Delta_3=\max[\frac{\gamma_{th2}}{(a_2-a_1\gamma_{th2})\rho},\frac{\gamma_{th1}}{a_1\rho} ]$, $\Delta_2=\max[\lambda_1,\frac{\gamma_{th1}(1-\alpha)}{a_1\alpha2\mu \rho}]$.\\ 
		\textit{Proof:} See Appendix C
			\vspace{-3mm}
		\subsection{Outage probability of $U_2$ under perfect CSI}
			The outage probability of $U_2$ under  perfect CSI is obtained by considering the perfect CSI assumption as used in $P_{out,1}^{P}$. 
			Thus the outage probability of $U_2$ under perfect CSI is given as 	\vspace{-3mm}
				\begin{align}\label{P_F}
	    P_{out,2}^{P}=&1-B_2 \nonumber\left(\beta_{r_k,2}\lambda_1\right)^n\Big[\iota_1 \left( \beta_{r_k,2}\lambda_1\right)^{l_1}\nonumber \nonumber\left(\beta_{s,r_k}\right)^{n+l_1}\\&\Gamma\left(\kappa_3,\chi\right)\nonumber+A_2
	    \nonumber\iota_2\left(k+1\right)^{\kappa_4}\nonumber\left( \beta_{r_k,2}\lambda_1\right)^{l_2}\\&\left(\beta_{s,r_k}\right)^{n+l_2-\Bar{i}}\Gamma\left(\kappa_4,(k+1)\chi\right)\Big],
	\end{align}
		where $B_2=\frac{K}{(m_{s,r_k}-1)!}\sum_{n=0}^{m_{r_k,2}-1}\frac{1}{n!}$, $\chi=\frac{m_{s,r_k}\gamma_{th2}}{\lambda_{s,r_k}(a_2-a_1\gamma_{th2})\rho} $. Further, when  $\gamma_{th2}>\frac{a_2}{a_1}$,  $P^P_{out,2}=1$. 	\vspace{-5mm}
  \section{Asymptotic Outage Probability}
    In this section, an approximation of the outage probability at a high SNR region is provided. The asymptotic outage probability is obtained by considering the $\rho \to \infty$. Further, at high SNR, the approximated CDF is given as $F(x)\underset{x\to 0}{\approx} \frac{1}{m_{i,j}!}\left(\beta_{i,j}x\right)^{m_{i,j}}$ \cite{jointharcsi}. 
     	\vspace{-4mm}
    \subsection{Asymptotic outage probability of $U_1$ under imperfect CSI and imperfect SIC}
    At high SNR, the asymptotic outage probability of $U_1$ is obtained by utilizing the approximated CDF equation in \eqref{e} and  equating $\frac{1}{\rho}\to 0$ in $\lambda_1$ and $\lambda_2$. The asymptotic outage probability of $U_1$ under imperfect CSI/SIC is given as
\begin{align} 
\small
    P_{out,1}^{asy}=&1-\bigg[\left(1-\frac{1}{m_{s,r_k}!}\left(\hat{\beta}_{s,r_k}\psi\right)^{m_{s,r_k}}\right)^K \nonumber\\&e^{-\hat{\beta}_{r_k,1}a}\sum_{n=0}^{m_{r_K,1}-1}\frac{1}{n!}\left(\hat{\beta}_{r_k,1}a\right)^n\bigg],
\end{align}
where, $a=\max[\Omega_1,\Omega_2]$. 	\vspace{-5mm}
\subsection{Asymptotic outage probability of $U_2$ under imperfect CSI}
The asymptotic outage probability of $U_2$ is obtained by  approximating CDF equation and  equating $\frac{1}{\rho}\to 0$ in $\lambda_1$. The asymptotic outage probability of $U_2$  under imperfect CSI is given as 	\vspace{-3mm}
\begin{align}
\small
    P_{out,2}^{asy}=&1-\bigg[\left(1-\frac{1}{m_{s,r_k}}\left(\hat{\beta}_{s,r_k}\Delta_1\right)^{m_{s,r_k}}\right)^K\nonumber\\& e^{-\hat{\beta}_{r_k,2}\Omega_1}\sum_{n=0}^{m_{r_K,2}-1}\frac{1}{n!}\left(\hat{\beta}_{r_k,2}\Omega_1\right)^n\bigg].
\end{align}\vspace{-6mm}
\subsection{Asymptotic outage probability of $U_1$ under perfect CSI and perfect SIC}
The asymptotic outage probability  of $U_1$ under perfect CSI and SIC is obtained by utilizing the approximated CDF. The asymptotic outage probability of $U_1$ under perfect CSI/SIC  is given by
	\begin{align}
	\small
	    P_{out,1}^{P,asy}=&\left(\frac{1}{m_{s,r_k}!}\left(\beta_{s,r_k}\Delta_3\right)^{m_{s,r_k}}\right)^K+\nonumber\frac{\left(\beta_{r_k,1}\Delta_2\right)^{m_{r_k,1}}}{ (m_{s,r_k}-1)!}\nonumber\\&\frac{K}{m_{r_k,1}!}\nonumber\big[ \nonumber\beta_{s,r_k}^{m_{r_k,1}}\nonumber\Gamma\left(\kappa_5,\beta_{s,r_k}\Delta_3\right)\nonumber
	+A_2\nonumber \beta_{s,r_k}^{(m_{r_k,1}-\Bar{i})}\end{align}\begin{align}
	&\left(k+1\right)^{-(\kappa_5+\Bar{i)}}\Gamma\left(\kappa_5+\Bar{i},(k+1)\beta_{s,r_k}\Delta_1\right)\big].
	\end{align}
	where $\kappa_5=m_{s,r_k}-m_{r_k,1}$.
	\vspace{-3mm}
\subsection{Asymptotic outage probability of $U_2$ under  perfect CSI}
 The asymptotic outage probability  of $U_2$ under perfect CSI is obtained by approximating CDF. The asymptotic outage probability of $U_2$ under perfect CSI  is given by
		\begin{align}
		\small
	    P_{out,2}^{P,asy}=&\left(\frac{1}{m_{s,r_k}!}
	    \chi^{m_{s,r_k}}\right)^K\nonumber+\nonumber\frac{K\left(\beta_{r_k,2}\lambda_1\right)^{m_{r_k,2}}}{(m_{s,r_k}-1)!m_{r_k,2}!}\nonumber\\&\nonumber\big[ \nonumber\beta_{s,r_k}^{m_{
	    r_k,2}}\nonumber\Gamma\left(\kappa_6,\chi\right)\nonumber+A_2
	    \nonumber \left(k+1\right)^{-\left(\kappa_6+i\right)}\\&\left(\beta_{s,r_k}\right)^{m_{r_k,2}-\Bar{i}} \Gamma\left(\kappa_6+\Bar{i},(k+1)\chi \right)\big].
	\end{align}
		where $\kappa_6=m_{s,r_k}-m_{r_k,2}$. 	\vspace{-3mm}
	 \subsection{Optimization of fraction of EH time ($\alpha$)}
	In this section, fraction of EH time ($\alpha$) is optimized in order to maximize the throughput of the system. The system throughput of a dual-hop system in a delay-limited transmission mode for a fixed transmission rate from the outage probability is given as \cite{shaik2020}
		\begin{align}\label{thrput}
		\tau=\frac{(1-P_{out,1})r_{1}(1-\alpha)}{2}+\frac{(1-P_{out,2})r_{2}(1-\alpha)}{2},
	\end{align} 
	where $\lambda_{th1}=2^{\frac{2r_{1}}{1-\alpha}}-1$ and $\lambda_{th2}=2^{\frac{2r_{2}}{1-\alpha}}-1$ are the threshold SNR for a fixed rate $r_{1}$ and $r_2$ respectively. Maximum throughput can be attained by optimizing $\alpha$. The objective function for maximizing throughput can be formulated as follows
	\begin{equation}\label{opt_alp}
		\begin{aligned}
			&\alpha^{*}= \text{arg} \max_{\alpha} {\tau}  \quad  \textrm{subject to} \quad 0<\alpha<1, \\
			&\alpha^{*}= \text{arg} \max_{\alpha} \quad  \frac{(1-P_{out,1})R_{1}(1-\alpha)}{2}\\ &\quad \quad \quad \quad \quad \quad+\frac{(1-P_{out,2})R_{2}(1-\alpha)}{2}	\\&
			\hspace{2.2em} \textrm{subject to} \quad 0<\alpha<1 	
		\end{aligned}
	\end{equation}
 The objective function in \eqref{opt_alp} is nonlinear and non-convex. Thus, we propose a low-complexity optimum fraction of EH time that maximize the system throughput based on a  PSO algorithm \cite{PSO}, wherein the optimal solution from each iteration in the search space is based on swarm of particles. It is noteworthy that we choose PSO  because it offers fast convergence and  stability \cite{PSO_lett, PSO_access}.  
The algorithm for determining the PSO-based solution is shown in Algorithm 1. Let $\tau(\alpha)$ denote the objective value of solution $\alpha$ as given in \eqref{thrput}. Let $\alpha_i$ denotes the position of particle $i$ ($1\leq i \leq MAX_{particles}$), where $MAX_{particles}$ denotes the number of particles.
\begin{algorithm}
\SetAlgoLined
 \KwIn{ Lower Bound of Decision Variables $=0$, Upper Bound of Decision Variables $=1$,}
 \KwOut{the best fitness value (optPosition)  and the corresponding solution (optcost)}
 Initialize $GlobalBest = -Infinity$
 \For{each particle i $\leq $ $MAX_{particles}$}
 {initialize $\alpha_i$, in the interval of [$0$, $1$] randomly 
 \\initialize velocity, $v_i=0$\\
 Compute the fitness value of particle $i$, $\tau(\alpha_i)$ and set the best solution of particle $i$ as $BestCost_i$ and corresponding position $BestPosition_i$.
 \If{the  $BestCost_i$ is greater than the $GlobalBest$}{
   set current value as the new $GlobalBest$ value and corresponding position $GlobalPosition$}}
   \For{  t $\leq$ 20}{
   \For{each particle i $\leq$  $MAX_{particles}$}{calculate the velocity of particle $i$\\$v_i=  v_i  + 2*$random function in the interval of [$0$, $1$]$*(BestPosition_i$ \\$- \alpha_i)+ 2*$random function in the interval of [$0$, $1$]$*(GlobalPosition - \alpha_i)$ \\update the particle position \\$\alpha_i = \alpha_i + v_i;$\\ 
    $\alpha_i = max(\alpha_i, 0);$\\
    $\alpha_i = min(\alpha_i, 1)$\\
    calculate the fitness value $\tau(\alpha_i)$ according to new position
    \If{the fitness value  is greater than the best fitness
value in history}{set current value as the new $BestCost_i$ of particle} 
    \If{the fitness value  is greater than the global best
value in history}{set current value as the new $GlobalBest$ value and corresponding position $GlobalPosition$}}
set optcost= $GlobalBest$ and optPosition= $GlobalPosition$}
 \caption{PSO algorithms}
\end{algorithm}
	\section{Ergodic Rate}
    In this section, we derive the ergodic rate expression for NOMA users. The  achievable  rate for error-free transmission is given as
    \begin{align}\label{e0}
        R_i&=E\left[\frac{1-\alpha}{2}\log_2(1+\gamma_{i,j})\right],\nonumber\\
        &=\frac{1-\alpha}{2\ln2}\int_{z=0}^{\infty}\frac{1-F_{i,j}(z)}{1+z}dz.
    \end{align}
    where $F_{i,j}$ represents CDF of SNR $\gamma_{i,j}$.
    The closed-form ergodic rate expressions for $U_1$ and $U_2$ are obtained in the following subsections. 	\vspace{-4mm}
	\subsection{Ergodic rate of $U_1$ under imperfect CSI and imperfect SIC}
     The ergodic rate of $U_1$ is defined as
     \begin{align}
         R_1=& E\left[\frac{1-\alpha}{2}\log_2\left(1+\min(\gamma_{r_k,1},\gamma_{1,1})\right)\right],\label{ern}\\
         R_1=& E\bigg[\frac{1-\alpha}{2}\log_2(1+\min(\nonumber\\&\frac{|\hat{h}_{s,r}|^2P_sa_1}{|\hat{h}_{s,r}|^2 \beta P_s a_2 +P_s\sigma^2_{e}( \beta a_2+a_1)  +\sigma^2_0},\nonumber\\& \frac{|\hat{h}_{r_k,1}|^2P_ra_1}{|\hat{h}_{r_k,1}|^2 \beta P_r a_2 +P_r\sigma^2_{e}( \beta a_2+a_1)  +\sigma^2_0})\bigg].\label{e1}
     \end{align}\vspace{-1mm}
Considering the practical scenario, the harvested energy at the relay is always small. Hence the transmit power of the relay is much lower than that of the source. Thus it is assumed that the SINR at the destinations is lower than the SINR at the relay \cite{PerEHOE}.
 \begin{align}
     &\frac{|\hat{h}_{s,r}|^2P_sa_1}{|\hat{h}_{s,r}|^2 \beta P_s a_2 +P_s\sigma^2_{e}( \beta a_2+a_1)  +\sigma^2_0}\nonumber>\\& \frac{|\hat{h}_{r_k,1}|^2P_ra_1}{|\hat{h}_{r_k,1}|^2 \beta P_r a_2 +P_r\sigma^2_{e}( \beta a_2+a_1)  +\sigma^2_0}.
 \end{align}
 Therefore, \eqref{e1} will reduce to  
  \begin{align}\label{26}
         R_1=& E\bigg[\frac{1-\alpha}{2}\log_2\Big(1+\nonumber\\&\frac{|\hat{h}_{r_k,1}|^2P_ra_1}{|\hat{h}_{r_k,1}|^2 \beta P_r a_2 +P_r\sigma^2_{e}( \beta a_2+a_1)  +\sigma^2_0}\Big)\bigg],
     \end{align}
     After some manipulation and by using \eqref{e0}, the above equation can be further simplified as
          \begin{align}\label{27}
          R_1=& \frac{1-\alpha}{2\ln2}\int_{z=0}^{\infty}\frac{1-F_{P}(z)}{1+z}dz\nonumber\\&
          -\frac{1-\alpha}{2\ln2}\int_{z=0}^{\infty}\frac{1-F_{Q}(z)}{1+z}dz,
     \end{align}
     where $P=|\hat{h}_{r_k,1}|^2 P_r(\beta a_2+a_1) +P_r\sigma^2_{e}( \beta a_2+a_1)$ and $Q=|\hat{h}_{r_k,1}|^2 \beta P_r a_2 +P_r\sigma^2_{e}( \beta a_2+a_1) $.  
     After some manipulation and using \cite[eq.(3.471.9)]{gradshteyn2014table}, the CDF of P and Q are given as follow
     \begin{align}\label{P}
     \small
         F_P(z)=&1-2e^{\hat{\beta}_{r_k,1}\sigma_e^2}A_1\nonumber(-\sigma_{e}^2)^{n-j}  \hat{\beta}_{s,r_k}^{m_{s,r_k}}\nonumber
         \left(\frac{z}{\zeta }\right)^j\\&\Bigg[\Big(\frac{\hat{\beta}_{r_k,1}z}{\hat{\beta}_{s,r_k}\zeta}\Big)^{\frac{\tau_1}{2}}\nonumber K_{\tau_1}\left(2\sqrt{\xi_1 z}\right)\nonumber
        +A_2\Big(\frac{\hat{\beta}_{r_k,1}}{\hat{\beta}_{s,r_k}}\Big)^{\frac{\tau_1+\Bar{i}}{2}}\\& \left(\frac{z}{\zeta(k+1)}\right)^{\frac{\tau_1+\Bar{i}}{2}}K_{\tau_1+\Bar{i}}\left(2\sqrt{\xi_1z(k+1)}\right)\Bigg],
     \end{align}
     where  $\tau_1=m_{s,r_k}-j$, $\xi_1=\frac{\hat{\beta}_{r_k,1}\hat{\beta}_{s,r_k}}{\zeta}$ and $\zeta=\frac{2\rho\mu \alpha (a_1+\beta a_2)}{1-\alpha}$. 	\vspace{-3mm}
     \begin{align}\label{Q}
     \small
         F_Q(z)=&1-2e^{\hat{\beta}_{r_k,1}\xi_3}\nonumber A_1 (-\xi_3)^{n-j}  \hat{\beta}_{s,r_k}^{m_{s,r_k}}\nonumber \left(\frac{z}{\Bar{\zeta} }\right)^j\\&\Big[\nonumber K_{\tau_1}\left(2\sqrt{\xi_2 z}\right)\nonumber\Big(\frac{\hat{\beta}_{r_k,1}z}{\hat{\beta}_{s,r_k}\Bar{\zeta}}\Big)^{\frac{\tau_1}{2}}+A_2\Big(\frac{\hat{\beta}_{r_k,1}}{\hat{\beta}_{s,r_k}}\Big)^{\frac{\tau_1+\Bar{i}}{2}}\\& \left(\frac{z}{\Bar{\zeta}(k+1)}\right)^{\frac{\tau_1+\Bar{i}}{2}}\ K_{\tau_1+\Bar{i}}\left(2\sqrt{\xi_2z(k+1)}\right)\Big],
     \end{align}
      where $\xi_2=\frac{\hat{\beta}_{r_k,1}\hat{\beta}_{s,r_k}}{\Bar{\zeta}}$, $\xi_3=\frac{\sigma_{e}^2 (a_1+\beta a_2)}{ \beta a_2}$ and $\Bar{\zeta}=\frac{2\rho\mu \alpha \beta a_2}{1-\alpha}$. Further, substituting \eqref{P} and \eqref{Q}  in \eqref{27}, and using \cite[ eq. (7.811.5), (9.34.3)]{gradshteyn2014table}, after some manipulation the ergodic rate of $U_1$ is given as 
\begin{align}
    R_1=\frac{1-\alpha}{2\ln2}I_1
\end{align}
where $I_1$ is defined in \eqref{i1}.
\vspace{-3mm}
\subsection{Ergodic rate of $U_2$ under imperfect CSI}
The ergodic rate of $U_2$ is define as
 \begin{align}\label{r_f}
 \small
         R_2=& E\left[\frac{1-\alpha}{2}\log_2\left(1+\min(\gamma_{r_k,2},\gamma_{2,2})\right)\right],
     \end{align}
 Substituting \eqref{rf} and \eqref{eqC} in \eqref{r_f}, and by employing the same approach as of $U_1$, the ergodic rate of $U_2$ is given by 
\begin{align}
\small
    R_2=\frac{1-\alpha}{2\ln2}J_1, 
\end{align}
where $J_1$ is defined in \eqref{j1}.
where $\nu=\frac{2\rho\mu \alpha }{1-\alpha}$ and $\Bar{\nu}=\frac{2\rho a_1\mu \alpha }{1-\alpha}$.\vspace{-3mm}
\subsection{Ergodic rate of $U_1$ under perfect CSI and perfect SIC}
		The ergodic rate of $U_1$ is obtained by substituting the SINR under the assumption of perfect CSI/SIC in \eqref{ern}. The ergodic rate of $U_1$ under perfect CSI/SIC is given as
		\begin{align}
		\small
	   R_1^P=&\frac{1-\alpha}{2\ln2}B_1\nonumber
         \Big[\nonumber  \MeijerG{1}{3}{3}{1}{0}{0,m_{s,r_k},n}{\xi_4}
         +A_{2} \beta_{s,r_k}^{-\Bar{i}}\\&\left(\frac{1}{k+1}\right)^{m_{s,r_k}+\Bar{i}}   \MeijerG{1}{3}{3}{1}{0}{0,m_{s,r_k}+\Bar{i},n}{\xi_4}\Big].
	\end{align}
	where $\xi_4=\frac{\beta_{s,r_k}\beta_{r_k,1}}{\Bar{\nu}}$. \vspace{-2mm}
		\subsection{Ergodic rate of $U_2$ under  perfect CSI}
			The ergodic rate of $U_2$ is obtained by substituting the SINR under the perfect CSI assumption in \eqref{r_f}.	The ergodic rate of $U_2$ under perfect CSI is given as
				\begin{align}
				\small
	   R_2^P=&\frac{1-\alpha}{2\ln2}\bigg[B_2\nonumber
         \Bigg(\nonumber  \MeijerG{1}{3}{3}{1}{0}{0,m_{s,r_k},n}{\xi_5}         +A_2 \beta_{s,r_k}^{-\Bar{i}}\\&\Big(\frac{1}{k+1}\Big)^{m_{s,r_k}+\Bar{i}}\nonumber    \MeijerG{1}{3}{3}{1}{0}{0,m_{s,r_k}+\Bar{i},n}{\xi_5 (k+1)}\bigg)\nonumber\\-&B_2\nonumber
         \Bigg(\nonumber  \MeijerG{1}{3}{3}{1}{0}{0,m_{s,r_k},n}{\xi_6} \nonumber +A_2\Big(\frac{1}{k+1}\Big)^{m_{s,r_k}+\Bar{i}}    \\&\beta_{s,r_k}^{-\Bar{i}} \MeijerG{1}{3}{3}{1}{0}{0,m_{s,r_k}+\Bar{i},n}{(k+1)\xi_6}\Bigg)\Bigg],
	\end{align}
	where $\xi_5=\frac{\beta_{s,r_k}\beta_{r_k,2}}{\nu}$ and $\xi_6=\frac{\beta_{s,r_k}\beta_{r_k,2}}{\Bar{\nu}}$.
	\vspace{-3mm}
	\section{Numerical and Simulation Results}
\begin{figure}.
    \centering
    \includegraphics[scale=0.59]{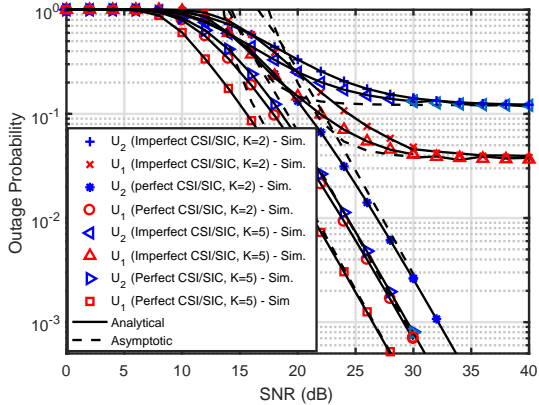}
    \caption{Outage probability of $U_1$ and $U_2$ with respect to transmit SNR ( $a_2=0.7$, $\alpha=0.35$).}
    \label{fig:1}
\end{figure}

	In this section, numerical and simulation results are presented to evaluate the impact of imperfect CSI/SIC, the number of relays, EH time fraction ($\alpha$) and $\sigma_e^2$ on the considered NOMA system. Unless specified, the system parameter are as follows. The target data rate $r_2=r_1=0.5$ bpcu, $\alpha=0.35$, energy conversion efficiency $\mu=0.9$, $m_{i,j}=2$, $\beta=0.15$, the channel gains $\lambda_{s,r_k}=\lambda_{r_k,1}=1$,  $\lambda_{r_k,2}=0.5$, $\sigma_{e}^2=0.01$, $a_1=0.3$ and $MAX_{particles}=30$.  	 \cite{PerEHOE}.  The correctness of the derived analytical expressions is validated through Monte-Carlo simulations. Simulations are performed using Matlab, and analytical results are obtained using Mathematica. In the figures, (Sim.) denotes Matlab simulation result.\\
		In \figurename{ \ref{fig:1}}, the outage probability experienced by $U_1$ and $U_2$ under imperfect CSI/SIC is compared with perfect CSI/SIC for the considered NOMA system. Results show a significant impact on the outage probability of both users due to imperfect CSI/SIC.  It is observed that at the outage of $10^{-1}$ with K=2,  $U_1$ with perfect CSI/SIC provides SNR gain of 5 dB over imperfect CSI/SIC case and $U_2$ with perfect CSI/SIC provides SNR gain of 10 dB over imperfect CSI/SIC.  Whereas, at high SNR regime, the outage probability of users with imperfect CSI/SIC suffers outage floor and reaches a constant value of 0.038 and 0.12 for $U_1$ and $U_2$ respectively. With the increase in SNR, CEE increases  and thus limits outage probability to further decrease and maintains a constant value for both users under imperfect CSI. However, in $U_1$, the residual interference due to imperfect SIC also increases with the increase in the SNR, thus causing degradation in the outage probability of $U_1$.  Further, from the figure, a performance gain is observed as we increase the K. A gain of 3 dB is observed for both users under perfect CSI/SIC, with the increase in relays from 2 to 5 for an outage probability of  $10^{-2}$.

\begin{strip}
\noindent\rule{\textwidth}{0.1pt}
\begin{align}\label{i1}
\small         I_1=& e^{\hat{\beta}_{r_k,1}\sigma_e^2}\nonumber
        A_1(-\sigma_{e}^2)^{n-j} \nonumber\left(\frac{1}{\zeta }\right)^j
         \Big[\left(\frac{\hat{\beta}_{r_k,1}}{\zeta}\right)^{-j}\nonumber  \MeijerG{1}{3}{3}{1}{0}{0,m_{s,r_k},j}{\xi_1}
         +A_2   \left(\frac{\hat{\beta}_{r_k,1}}{\zeta}\right)^{-j}\nonumber \left(\frac{1}{k+1}\right)^{m_{s,r_k}+\Bar{i}}  \hat{\beta}_{s,r_k}^{-\Bar{i}}
        \\&\MeijerG{1}{3}{3}{1}{0}{0,m_{s,r_k}+\Bar{i},j}{\xi_1(k+1)}\Big]
         - e^{\xi_3\hat{\beta}_{r_k,1}}A_1
        \left(-\xi_3\right)^{n-j} \nonumber\left(\frac{1}{\Bar{\zeta} }\right)^j\nonumber
         \Big[\left(\frac{\hat{\beta}_{r_k,1}}{\Bar{\zeta}}\right)^{-j}\nonumber  \MeijerG{1}{3}{3}{1}{0}{0,m_{s,r_k},j}{\xi_2}\
         +A_2\\& \left(\frac{\hat{\beta}_{r_k,1}}{\Bar{\zeta}}\right)^{-j}\left(\frac{1}{k+1}\right)^{m_{s,r_k}+\Bar{i}} \hat{\beta}_{s,r_k}^{-\Bar{i}} \MeijerG{1}{3}{3}{1}{0}{0,m_{s,r_k}+\Bar{i},j}{\xi_2(k+1)}\Big].\\
         J_1=& e^{\hat{\beta}_{r_k,2}\sigma_e^2}\nonumber A_3
        (-\sigma_{e}^2)^{n-j} \nonumber\left(\frac{1}{\nu }\right)^j
         \Big[\left(\frac{\hat{\beta}_{r_k,2}}{\nu}\right)^{-j}\nonumber  \MeijerG{1}{3}{3}{1}{0}{0,m_{s,r_k},j}{\xi_5}
         +A_2  \nonumber \left(\frac{\hat{\beta}_{r_k,2}}{\nu}\right)^{-j}\left(\frac{1}{k+1}\right)^{m_{s,r_k}+\Bar{i}} \hat{\beta}_{s,r_k}^{-\Bar{i}}\\&\nonumber\MeijerG{1}{3}{3}{1}{0}{0,m_{s,r_k}+\Bar{i},j}{\xi_5(k+1)}\Big]- e^{\frac{\hat{\beta}_{r_k,2}\sigma_e^2 }{  a_1}}A_3\nonumber
        \left(\frac{-\sigma_{e}^2 }{  a_1}\right)^{n-j} \nonumber\left(\frac{1}{\Bar{\nu} }\right)^j
         \Big[\nonumber  \MeijerG{1}{3}{3}{1}{0}{0,m_{s,r_k},j}{\xi_6}\left(\frac{\hat{\beta}_{r_k,2}}{\Bar{\nu}}\right)^{-j}
         \\&+A_2   \left(\frac{\hat{\beta}_{r_k,2}}{\Bar{\nu}}\right)^{-j}\left(\frac{1}{k+1}\right)^{m_{s,r_k}+\Bar{i}} \hat{\beta}_{s,r_k}^{-\Bar{i}}\MeijerG{1}{3}{3}{1}{0}{0,m_{s,r_k}+\Bar{i},j}{\xi_6(k+1)}\Big].\label{j1}
\end{align}
\noindent\rule{\textwidth}{0.4pt}
\end{strip}

\begin{figure}
     \centering
     \includegraphics[scale=0.6]{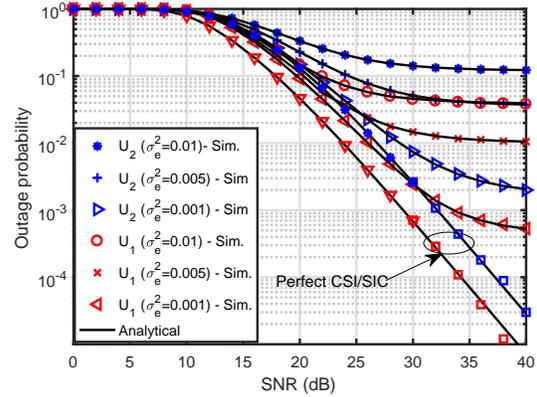}
     \caption{Outage probability of $U_1$ and $U_2$ under imperfect CSI/SIC  with respect to transmit SNR  for different value of $\sigma_e^2$ ( $a_2=0.7$, $\alpha=0.35$).}
     \label{fig:3}
 \end{figure}

\begin{figure}
     \centering
     \includegraphics[scale=0.6]{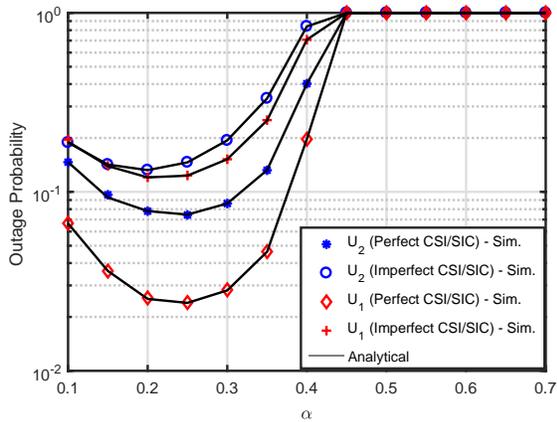}
     \caption{Outage probability of $U_1$ and $U_2$ with respect to $\alpha$ (K=2, $\rho=20 dB$, $a_2=0.7$)}
     \label{fig:4}
 \end{figure}

\begin{figure}
     \centering
     \includegraphics[scale=0.6]{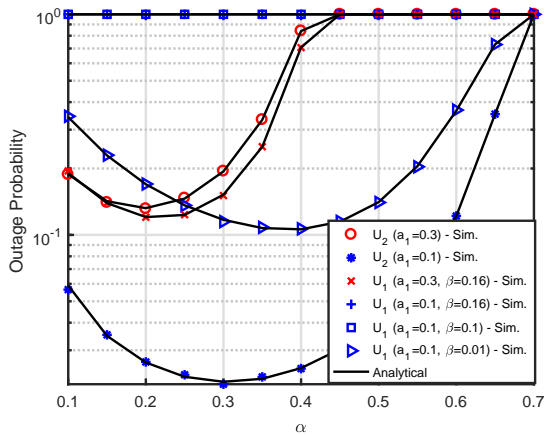}
     \caption{Outage probability of $U_1$ and $U_2$ under imperfect CSI/SIC  with respect to $\alpha$ (K=2, $\rho=20 dB$, $\sigma_e^2=0.01$)}
     \label{fig:5}
 \end{figure}

  In the case of imperfect CSI/SIC, an SNR gain of 2 dB is observed at an outage probability of 0.1 and 0.2 for $U_1$ and $U_2$, respectively. Whereas at high value of SNR, due to the presence of CEE and residual interference due to imperfect SCI performance gain is not observed with the increase in K. It is also observed that the analytical results are perfectly matching with the simulation result.  Further, the derived asymptotic outage results match the derived analytical and the simulation results at a high SNR value, which validates our results.	 \\
\figurename{ \ref{fig:3}} investigates the impact of CEE $\sigma^2_e$ on  $U_1$ and $U_2$ with the fixed residual interference due to SIC error $\beta$. It is observed the outage floor decreases with a decrease in $\sigma^2_e$ and approaches towards perfect CSI/SIC case for both the users. In $U_1$, for high CEE values $\sigma^2_e = 0.01$ and $\sigma^2_e= 0.005$, perfect CSI/SIC has an SNR gain of 4 dB and 3 dB for an outage of $10^{-1}$. However, for smaller values of CEE $(\sigma^2_e = 0.001)$, the gain is 5 dB for an outage of $10^{-3}$. Thus, the outage probability is more limited by CEE than SIC error, as high CEE shows degradation in outage probability with the same value of $\beta$.\\
 %
\figurename{ \ref{fig:4}} and \figurename{ \ref{fig:5}} illustrate the impact of $\alpha$, the fraction of EH time, on the outage probability of $U_1$ and $U_2$. In the given scenario, the $\alpha$ is ranged between 0.1 and 0.7, whereas the $\rho$ and $\sigma_e^2$ are fixed at 20 dB and 0.01 respectively. In \figurename{ \ref{fig:4}}, the system's performance comparison under perfect CSI/SIC and imperfect CSI/SIC are shown.  With the increase in $\alpha$, the EH time increases, thus reducing the information processing time and hence outage probability of users increases for the entire time duration. The plot shows that the optimum value of $\alpha$, which minimizes the outage probability, lies in the range of 0.2 to 0.3 for both perfect and imperfect CSI/SIC.  Further, it is observed that at a high value of $\alpha$, i.e., above 0.45, the outage probability of users become one. With the increase in the $\alpha$, the threshold SNR, i.e., $\gamma_{th1}$ and $\gamma_{th2}$, also increases to maintain the constant target data rates $r_1$ and $r_2$, respectively. Thus, the outage probability criteria $\gamma_{th2}<\frac{a_2}{a_1}$ and $\gamma_{th1}<\frac{a_1}{\beta a_2}$ does not satisfied and makes the outage probability $P_{out,1}=P_{out,2}=1$. In \figurename{ \ref{fig:5}}, the impact of  $a_1$ and $\beta$ on the outage probability of users under imperfect CSI/SIC is analyzed. In the case of $U_2$, it is observed that the outage probability with $a_1=0.1$ shows significant improvement over $a_1=0.3$ since more power is allocated to the symbol intended for $U_2$. However, in $U_1$, $P_{out,1}$ tends to unity due to failure of outage criteria $\gamma_{th1}<\frac{a_1}{\beta a_2}$ and $\gamma_{th2}<\frac{a_2}{ a_1}$.  Considering the high residual interference due to imperfect SIC, i.e., $\beta=0.16$ and $\beta=0.1$, the outage criteria is not satisfied and leads to constant noise error floor. With a further decrease in residual interference i.e. $\beta=0.01$, the outage probability of $U_1$ reduces. Hence, selection of $a_1$ and $\beta$ plays a crucial role on the outage performance of $U_1$.    \\       
  \begin{figure}
     \centering
     \includegraphics[scale=0.61]{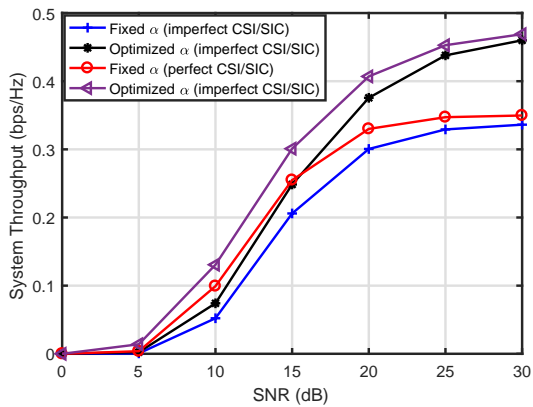}
     \caption{System throughput with respect to transmit SNR, comparison study of optimized and non optimized fraction of EH time (K=2, $a_1=0.3$)}
     \label{fig:7}
 \end{figure}
 \begin{figure}
     \centering
     \includegraphics[scale=0.59]{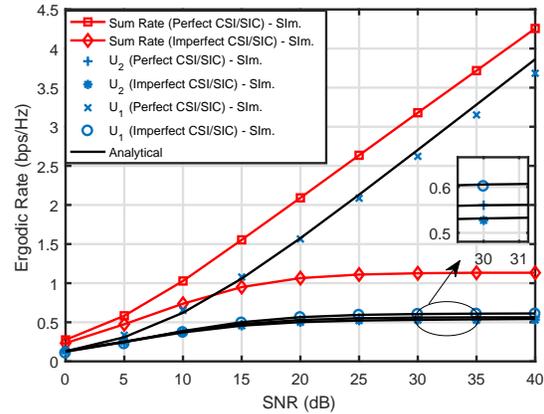}
     \caption{Ergidic rate of $U_1$ and $U_2$  with respect to transmit SNR (K=2, $a_1=0.3$)}
     \label{fig:6}
 \end{figure}
    \figurename{ \ref{fig:7}} shows the system throughput under optimized and non-optimized, i.e., (arbitrary)  $\alpha=0.3$ value. It is observed that  the system throughput of the optimized NOMA scheme under both perfect and imperfect SIC/CSI is enhanced compared to the non-optimized value. It is observed that at the system throughput of 0.3 bps/Hz, the optimized value provides an SNR gain of 3 dB over the non-optimized  fixed value.  
	In \figurename{ \ref{fig:6}},  both users' ergodic rate under perfect CSI/SIC and imperfect CSI/SIC are plotted against SNR. It is observed from the figure that the ergodic rate of $U_1$ under perfect CSI/SIC outperforms the ergodic rate of all other cases. The ergodic rate of $U_1$ under perfect CSI/SIC increases linearly with the SNR, whereas in case of $U_2$, the ergodic rate saturate at a high SNR region.	The ergodic rate of 0.5 bps/Hz the $U_1$ provides an SNR gain of 7 dB over $U_2$ with perfect CSI/SIC. With the increase in SNR, the interference generated by the symbol of $U_1$ increases and leads to saturation in the ergodic rate plot.	It is also observed that both users under imperfect CSI/SIC shows are marginal increasing trend at low SNR region and saturates at high SNR region. This is due to the increase in SNR, the interference due to imperfect CSI/SIC increases, thus the ergodic rate reduces effectively. Further, the NOMA system's sum rate (i.e., $r_1+r_2$) is also presented under perfect CSI/SIC and imperfect CSI/SIC. For the ergodic rate of 1 bps/Hz, the perfect CSI/SIC case provides an SNR gain of 10dB over imperfect CSI/SIC. It is observed that at high SNR values, there is a marginal gap between the analytical and simulated curve for $U_1$,  which is because of the approximation used in the analytical expression. 
	\vspace{-4mm}
\section{Conclusion}
In this paper, the analysis of downlink multiple EH relay-based  NOMA system over Nakagami$-m$ fading is performed. The practical assumption of imperfect SIC and imperfect CSI were taken into consideration for investigation.  The closed-form expressions for the outage probability, asymptotic outage probability, and ergodic rate of the users under imperfect CSI/SIC are obtained. The analytical results showed that the system under imperfect CSI/SIC provides massive performance degradation compared to the perfect CSI/SIC NOMA system. It is observed that the users' performance improved with increasing the total number of active relay nodes in the system. The significance of channel estimation error over the performance of users is analyzed.  The impact of power allocation coefficient and a fraction of block time over the user performances are also analyzed. Further, the impact of SIC error over the performance of $U_1$ is analyzed. It is also observed that there is an optimum value of $\alpha$ for users' minimum outage probability.  Obtained analytical results are validated through extensive simulation results. This generalized analysis is useful for future applications, including UAV, THz, and Internet of Things deployments for beyond 5G networks.
 \vspace{-2mm}
\appendices
\numberwithin{equation}{section}
\section{Proof of lemma 1} 
\begin{proof}
    	The closed-from expression of \eqref{A} is derived as follows.
	Consider the outage probability definition given in \eqref{Pn}. Substituting \eqref{rf}, \eqref{rn}, \eqref{nf} and \eqref{nn} in \eqref{Pn}, on rearranging, we get
	%
		\begin{align}\label{c}
		\small
	 P_{out,1} =&1- \Pr\Big\{ |\hat{h}_{s,r}|^2 >\frac{ \gamma_{th2}(P_s\sigma^2_{e}+\sigma^2_0)}{(a_2-a_1\gamma_{th2}) P_s },\nonumber\\&
	 |\hat{h}_{r_k,1}|^2> \frac{\gamma_{th2}(P_r\sigma^2_{e}+\sigma^2_0)}{(a_2-a_1\gamma_{th2})P_r},\nonumber\\& 
	 |\hat{h}_{s,r}|^2>\frac{\gamma_{th1}( P_s \sigma^2_{e}(a_2+ \beta a_1)  +\sigma^2_0)}{(a_1-\beta a_2\gamma_{th1} )P_s},\nonumber\\&
	 |\hat{h}_{r_k,1}|^2
	 >\frac{\gamma_{th1}(P_r\sigma^2_{e}(a_1+\beta a_2) +\sigma^2_0)}{(a_1-\beta a_2\gamma_{th1} )P_r} \Big\},
	\end{align} 
 According to EH criteria the transmitted power at relay is given as $P_r=\frac{2 P_s\mu |\hat{h}_{s,r}|^2\alpha  }{1-\alpha}$. Substituting $P_r$ in \eqref{c}, and after some simplification \eqref{c} is given as
 	 \begin{align}\label{e}
 	 \small
	 P_{out,1} =&1- \Pr\Big\{ |\hat{h}_{s,r}|^2 >\psi,\,
	 |h_{r_k,1}|^2> \frac{\lambda_1}{|\hat{h}_{s,r}|^2}\nonumber\\& +\Omega_1, \,
	 |h_{r_k,1}|^2
	 >\frac{\lambda_2}{|\hat{h}_{s,r}|^2}+\Omega_2\Big\},
	\end{align}
	where, $\Delta_1=\frac{ \gamma_{th2}(\sigma^2_{e}+(1/\rho))}{(a_2-a_1\gamma_{th2})  }$, $\lambda_1=\frac{\gamma_{th2} \left(1-\alpha\right)}{\left(a_2-a_1\gamma_{th2}\right)\alpha\mu2\rho}$, 
	$\lambda_2=\frac{\gamma_{th1} \left(1-\alpha\right)}{\left(a_1-a_2\beta\gamma_{th1}\right)\alpha\mu2\rho}$,
	$\Omega_1=\frac{\gamma_{th2 }\sigma_e^2}{a_2-a_1\gamma_{th2}}$, $\Omega_2=\frac{\gamma_{th2}\left(a_1+a_2\beta\right) \sigma_e^2}{a_1-a_2\beta\gamma_{th2}}$,  $\psi=\max\left[\Delta_1, \frac{\gamma_{th1}\left( P_s \sigma_e^2\left(a_1+a_2\beta\right)+\sigma_0^2\right)}{\left(a_1-a_2\beta\gamma_{th1}\right)P_s} \right]$.
	 \begin{align}
	 \small
	 P_{out,1} 
	 =&1- \Big[\Pr\Big\{ |\hat{h}_{s,r}|^2 >\psi,\,
	 |h_{r_k,1}|^2> \frac{\lambda_1}{|\hat{h}_{s,r}|^2}\nonumber\\& +\Omega_1,\,
	|\hat{h}_{s,r}|^2 >\frac{\lambda_2-\lambda_1}{\Omega_1-\Omega_2}\Big\}\nonumber\\&
	 +\Pr\Big\{ |\hat{h}_{s,r}|^2 >\psi,\,
	 |h_{r_k,1}|^2> \frac{\lambda_2}{|\hat{h}_{s,r}|^2}\nonumber\\& +\Omega_2,\, |\hat{h}_{s,r}|^2
	 <\frac{\lambda_2-\lambda_1}{\Omega_1-\Omega_2}\Big\}\Big]\label{e3}.
	\end{align}
	Let $P_{out,1} =1- (A + B)$, where $A$ and $B$ are separately evaluated as follows. Assume the links in the network to be independent, $\gamma_{th1}<\frac{a_1}{a_2\beta}$,  $\gamma_{th2}<\frac{a_2}{a_1}$, $\lambda_2>\lambda_1$ and $\Omega_1>\Omega_2$. 
	 Since i.i.d Nakagami  fading is assumed, the channel power gains $|\hat{h}_{i,j}|$ are gamma function. The PDF and CDF \cite{goldsmith2005wireless} of $|\hat{h}_{i,j}|^2$ with parameter $m_{i,j}$ and $\hat{\lambda}_{i,j}$ is given as
	 \begin{equation}
	     f_{|\hat{h}_{i,j}|^2}(x)=\left(\frac{m_{i,j}}{\hat{\lambda}_{i,j}}\right)^{m_{i,j}}\frac{x^{m_{i,j}-1}}{\Gamma(m_{i,j})}e^{\frac{-x m_{i,j}}{\hat{\lambda}_{i,j}}},
	 \end{equation}
	 \begin{equation}
	     F_{|\hat{h}_{i,j}|^2}(x)=1-e^{\frac{-x m_{i,j}}{\hat{\lambda}_{i,j}}}\sum_{n=0}^{m_{i,j}-1}\frac{1}{n!}\left(\frac{x m_{i,j}}{\hat{\lambda}_{i,j}}\right)^n,
	 \end{equation}
	 Let $\hat{\beta}_{i,j}=\frac{m_{i,j}}{\hat{\lambda}_{i,j}}$, since PRS with K-relays is performed,the CDF and PDF of the $ |\hat{h}_{s,r}|^2$ is given by
	  \begin{equation}
	  \small
	     F_{|\hat{h}_{s,r}|^2}(x)=\left[1-e^{-x \hat{\beta}_{s,r_k}}\sum_{n=0}^{m_{s,r_k}-1}\frac{1}{n!}\left(x\hat{\beta}_{s,r_k}\right)^n\right]^K,
	 \end{equation}
	 \begin{align}\label{f}
	 \small
	     f_{|\hat{h}_{s,r}|^2}(x)=&K{\hat{\beta}_{s,r_k}}^{m_{s,r_k}}\frac{x^{m_{s,r_k}-1}}{(m_{s,r_k}-1)!}e^{-x\hat{\beta}_{s,r_k}}\nonumber\Bigg[1+\sum_{k=1}^{K-1}\\&(-1)^k \binom{K-1}{k}e^{-k x \hat{\beta}_{s,r_k}}
	     \nonumber\\&\left(\sum_{n=0}^{m_{s,r_k}-1}\frac{1}{n!}\left(x \hat{\beta}_{s,r_k} \right)^n\right)^k \Bigg],
	 \end{align}
	 By using the Binomial expression  for the term $\tiny \left(\sum_{n=0}^{m_{s,r_k}-1}\frac{1}{n!}\left(x \hat{\beta}_{s,r_k}\right)^n\right)^k$,	the Binomial expansion is given as \cite{ICSINOMA}
	 \begin{align}\label{g}
	    	 &=
	\sum_{i_1=0}^{k}\sum_{i_2=0}^{k-i_2}\dotsm\nonumber\sum_{i_{m_{s,r_k}-1}=0}^{k-i_1\cdots i_{m_{s,r_k}-2}}\nonumber\binom{k}{i_1}\binom{k-i_1}{i_2} \ldots\\& \binom{k-i_1-...i_{m_{s,r_k}-2}}{i_{m_{s,r_k}-1}}\nonumber \prod_{r=0}^{m_{s,r_k}-2}\left( \frac{{(\hat{\beta}_{s,r_k}x)}^r}{r!} \right)^{i_{r+1}} \nonumber\\&\left( \frac{\hat{\beta}_{s,r_k}^{m_{s,r_k}-1}x}{(m_{s,r_k}-1)!} \right)^{k-i_1-...i_{m_s,{r_k}-1}}  \end{align}
	Substituting \eqref{g} in \eqref{f}, the CDF is obtained. By utilizing the above equation A is obtained as follows
	\begin{align}\label{q}
	    A=&\int_{x=\gamma}^{\infty}\left[1-F_{r_k,1}\left(\frac{\lambda_1}{x}+\Omega_1\right)\right]f_{s,r_k}(x)dx,\end{align}\begin{align}
	    A=&\int_{x=\gamma}^{\infty}e^{-\hat{\beta}_{r_k,1}\left(\frac{\lambda_1}{x}+\Omega_1\right)}\sum_{n=0}^{m_{r_k,1}-1}\frac{K}{n!}\hat{\beta}_{s,r_k}^{m_{s,r_k}}e^{-x\hat{\beta}_{s,r_k}}\nonumber\\&\left(\hat{\beta}_{r_k,1}\left(\frac{\lambda_1}{x}+\Omega_1\right)\right)^n \nonumber\frac{x^{m_{s,r_k}-1}}{(m_{s,r_k}-1)!}\nonumber\Bigg[1+\sum_{k=1}^{K-1}\end{align}
	    \begin{align}
	    &(-1)^k \binom{K-1}{k}e^{-k x\hat{\beta}_{s,r_k}}\bigcup_k\Xi_{1,k}\Xi_{2,k}x^{\Bar{i}}\Bigg]dx,
	    \end{align}
	    where $\gamma=\max\left[\psi,\left(\frac{\lambda_2-\lambda_1}{\Omega_1-\Omega_2}\right)\right], \, \bigcup_k=\sum_{i_1=0}^{k}\sum_{i_2=0}^{k-i_2}\dotsm \sum_{i_{m_{s,r_k}=0
	    	}}^{k-i_1\cdots i_{m_{s,r_k}-2}}, \, \Xi_{1,k}= \binom{k}{i_1}\binom{k-i_1}{i_2} . . . \binom{k-i_1-...i_{m_{s,r_k}-2}}{i_{m_{s,r_k}-1}}, \, \Xi_{2,k}=\prod_{r=0}^{m_{s,r_k}-2}\left( \frac{\hat{\beta}_{s,r_k}^r}{r!} \right)^{i_{r+1}}  \left( \frac{\hat{\beta}_{s,r_k}^{m_{s,r_k}-1}}{(m_{s,r_k}-1)!} \right)^{k-i_1-...i_{m_s,{r_k}}-1}$ and $\Bar{i}= (m_{s,r_k}-1)(k-i_1)-(m_{s,r_k}-2)i_2-(m_{s,r_k}-3)i_3-...i_{m_{s,r_k}-1}$.
	    Using Binomial expansion on $\left(\frac{\lambda_1}{x}+\Omega_1\right)^n$, the above equation can further simplified as
	    \begin{align}\label{v}
	    A=&e^{-\hat{\beta}_{r_k,1}\Omega_1}\sum_{n=0}^{m_{r_k,1}-1}\frac{1}{n!}\hat{\beta}_{r_k,1}^n K\hat{\beta}_{s,r_k}^{m_{s,r_k}}\nonumber
	    \frac{1}{(m_{s,r_k}-1)!}\\&\sum_{j=0}^{n}\binom{n}{j}\lambda_1^j\Omega_1^{n-j}\Bigg[\int_{x=\gamma}^{\infty} \nonumber x^{m_{s,r_k}-j-1}
	    e^{-x\hat{\beta}_{s,r_k}}\\&e^{\frac{-\hat{\beta}_{r_k,1}\lambda_1}{x}}dx \nonumber+\sum_{k=1}^{K-1}(-1)^k \binom{K-1}{k}\bigcup_k\Xi_{1,k}
	   \Xi_{2,k}\\&\int_{x=\gamma}^{\infty}e^{-x(k+1)\hat{\beta}_{s,r_k}}x^{m_{s,r_k}-j+\Bar{i}-1}e^{\frac{- \hat{\beta}_{r_k,1}\lambda_1}{x}}dx\Bigg].
	\end{align}
	Now, using Taylor's series expansion, as  $e^{\frac{-\hat{\beta}_{r_k,1}\lambda_1}{x}}=\sum_{l_1=0}^{N_t}\frac{(-1)^{l_1}}{l_1!}\left(\frac{\hat{\beta}_{r_k,1}\lambda_1}{x}\right)^{l_1} $ where $N_t\in \{2, 3, ...\infty\}$ and using \cite[eq. (3.351.4)]{gradshteyn2014table} the \eqref{v} can be simplified.
	Further, the B is simplified as follows
	\begin{align}
	     B=&\int_{x=\psi}^{\frac{\lambda_2=\lambda_1}{\Omega_1-\Omega_2}}\left[1-F_{r_k,1}\left(\frac{\lambda_2}{x}+\Omega_2\right)\right]f_{s,r}(x)dx.\label{w2}
	\end{align}
	The \eqref{w2}, can be simplified by following the same approach as used in \eqref{q}, the obtained result $P_{out,1} =1- (A + B)$ is given by \eqref{A}. 
\end{proof}

\section{Proof of lemma 2}
\begin{proof}
The closed form expression of \eqref{P_F_I} is obtained as follows.
Substitute \eqref{rf} and \eqref{eqC} in \eqref{eq1}, after some manipulation \eqref{eq1} can be simplified as     
    	\begin{align}\label{f1}
	 P_{out,2} =&1- \Pr\Big\{ |\hat{h}_{s,r}|^2 >\frac{ \gamma_{th2}(P_s\sigma^2_{e}+\sigma^2_0)}{(a_2-a_1\gamma_{th2}) P_s },\nonumber\\&
	 |\hat{h}_{r_k,2}|^2> \frac{\gamma_{th2}(P_r\sigma^2_{e}+\sigma^2_0)}{(a_2-a_1\gamma_{th2})P_r}
	 \Big\},
	\end{align}
	Substituting $P_r=\frac{2P_s\alpha\mu|\hat{h}_{s,r}|^2}{1-\alpha}$, \eqref{f1} can be further reduced as
	 
		 \begin{align}\label{f2}
	 P_{out,2} =&1- \Pr\Big\{ |\hat{h}_{s,r}|^2 >\Delta_1,\nonumber\\&
	 |\hat{h}_{r_k,1}|^2> \frac{\lambda_1}{|\hat{h}_{s,r}|^2}+\Omega_1\Big\},
	\end{align} 
		\begin{align}\label{f3}
	    P_{out,2}=&1-\int_{x=\Delta_1}^{\infty}\left[1-F_{r_k,2}\left(\frac{\lambda_1}{x}+\Omega_1\right)\right]f_{s,r}(x)dx.
	    \end{align}
	    The integration in \eqref{f3} can be simply simplify by following the same approach as of \eqref{q} of Appendix A. The obtained closed-form expression is given by \eqref{P_F_I}.
\end{proof}

\section{Proof of lemma 3}
\begin{proof}
	Considering the  perfect CSI and perfect SIC condition on the SINR's equation \eqref{rf}, \eqref{rn}, \eqref{nf} and \eqref{nn}. The outage probability of $U_1$ under perfect SIC and perfect CSI is given as 
		\begin{align}\label{b1}
	 P_{out,1}^P =&1- \Pr\Big\{ \frac{|{h}_{s,r}|^2P_s a_2}{|{h}_{s,r}|^2 P_s a_1+\sigma^2_0}. \nonumber> \gamma_{th2},\\&
	 \frac{|{h}_{r_k,1}|^2P_r a_2}{|{h}_{r_k,1}|^2 P_r a_1+\sigma^2_0} > \gamma_{th2},\nonumber 
	 \frac{|{h}_{s,r}|^2P_sa_1}{\sigma^2_0}\\&>\gamma_{th1},
	 \frac{|{h}_{r_k,1}|^2P_ra_1}{\sigma^2_0}.
	 >\gamma_{th1}  \Big\}.
	\end{align} 
	substituting $P_r=\frac{2 P_s\mu |h_{s,r}|^2\alpha  }{1-\alpha}$ and  after some manipulation, \eqref{b1} further simplified as	
		\begin{align}\label{c1}
	 P_{out,1}^P =& 1- \Pr\Big\{ |{h}_{s,r}|^2 > \Delta_3,
	 |{h}_{r_k,1}|>\frac{\Delta_2}{|h_{s,r}|^2}\Big\},
	\end{align} 
	where $\Delta_3=\max[\frac{\gamma_{th2}\sigma_0^2}{(a_2-a_1\gamma_{th2})P_s},\frac{\gamma_{th1}\sigma_0^2}{a_1P_s} ]$, $\Delta_2=\max[\lambda_1,\frac{\gamma_{th1}(1-\alpha)}{a_1\alpha2\mu P_s}]$. Let $P_{out,1}^P=1-D$, where D is evaluated as follow 
	\begin{align}
	     D=&\int_{x=\Delta_3}^{\infty}\left[1-F_{r_k,1}\left(\frac{\Delta_2}{x}\right)\right]f_{s,r_k}(x)dx,\label{c4}
	       \\=&\sum_{n=0}^{m_{r_k,1}-1}    \frac{1}{(m_{s,r_k}-1)!}\left(\frac{ m_{r_k,1}\Delta_2}{\lambda_{r_k,1}}\right)^n \left(\frac{m_{s,r_k}}{\lambda_{s,r_k}}\right)^{m_{s,r_k}}\nonumber\end{align}
	       \begin{align}&\frac{K}{n!}\Bigg[\int_{x=\Delta_3}^{\infty} \nonumber x^{m_{s,r_k}-n-1}
	    e^{\frac{ m_{s,r_k}x}{\lambda_{s,r_k}}}e^{\frac{- m_{r_k,1}\Delta_2}{\lambda_{r_k,1}x}}dx \nonumber\\&+\sum_{k=1}^{K-1}(-1)^k \binom{K-1}{k}\bigcup_k\Xi_{1,k}\Xi_{2,k}\int_{x=\Delta_3}^{\infty}\nonumber\\&e^{\frac{-x(k+1) m_{s,r_k}}{\lambda_{s,r_k}}}\nonumber x^{m_{s,r_k}-n+\Bar{i}-1}e^{\frac{- m_{r_k,1}\Delta_2}{\lambda_{r_k,1}x}}dx\Bigg].\label{c2}
	\end{align}
	The integration in \eqref{c2}, is same as in \eqref{v} in appendix A. Hence, D can be obtained following the similar procedure as adopted in \eqref{v}, which is given by \eqref{p_N}. 
\end{proof}
	\bibliographystyle{IEEEtran}
	\bibliography{main}
	\end{document}